\documentclass[manuscript]{aastex}
\usepackage{mathtools}
\usepackage{epsfig}

\newcommand{\gr}{$\gamma$-ray}
\newcommand{\grs}{$\gamma$-rays}

\slugcomment{Not to appear in Nonlearned J., 45.}
\shorttitle{Observing \emph{Fermi} J1808.2-2029}
\shortauthors{Yeung et al.}

\begin{document}

\title{Studying the SGR 1806-20/Cl* 1806-20 region using the \emph{Fermi} Large Area Telescope}

\author{Paul K. H. Yeung\altaffilmark{1,2,3}}

\author{Albert K. H. Kong\altaffilmark{1}}

\author{P. H. Thomas Tam\altaffilmark{4}}

\author{Lupin C. C. Lin\altaffilmark{5}}

\author{C. Y. Hui\altaffilmark{3}}

\author{Chin-Ping Hu\altaffilmark{2}}

\and

\author{K. S. Cheng\altaffilmark{2}}

\altaffiltext{1}{Institute of Astronomy and Department of Physics, National Tsing Hua University, Hsinchu, Taiwan; \email{paul2012@connect.hku.hk}, \email{akong@phys.nthu.edu.tw}}
\altaffiltext{2}{Department of Physics, University of Hong Kong, Pokfulam Road, Hong Kong}
\altaffiltext{3}{Department of Astronomy and Space Science, Chungnam National University, Daejeon 305-764, Republic of Korea}
\altaffiltext{4}{Institute of Astronomy and Space Science, Sun Yat-Sen University, Guangzhou 510275, China; \email{tanbxuan@mail.sysu.edu.cn}}
\altaffiltext{5}{Institute of Astronomy and Astrophysics, Academia Sinica, Taiwan}

%% Notice that each of these authors has alternate affiliations, which
%% are identified by the \altaffilmark after each name.  Specify alternate
%% affiliation information with \altaffiltext, with one command per each
%% affiliation.

\begin{abstract}

The region around SGR 1806-20 and its host stellar cluster Cl* 1806-20 is a potentially important site of particle acceleration. The soft $\gamma-$ray repeater and  Cl* 1806-20, which also contains several very massive stars including a luminous blue variable hypergiant LBV 1806-20, are capable of depositing a large amount of energy to the surroundings. Using the data taken with the \emph{Fermi} Large Area Telescope (LAT), we identified an extended LAT source to the south-west of Cl* 1806-20. The centroid of the 1-50~GeV  emission  is consistent with that of HESS J1808-204 (until now unidentified). The LAT spectrum is best-fit by a broken power-law with the break energy  $E_\mathrm{b}=297\pm15$ MeV. The index above $E_\mathrm{b}$ is $2.60\pm0.04$, and is consistent with the flux and spectral index above 100 GeV for HESS J1808-204, suggesting an association between the two sources. Meanwhile, the interacting supernova remnant SNR G9.7-0.0 is also a potential contributor to the LAT flux. A tentative flux enhancement at the MeV band during a  45-day interval (2011 Jan 21 - 2011 Mar 7) is also reported. We discuss possible origins of the extended LAT source in the context of both leptonic and hadronic scenarios.

\end{abstract}


\section{Introduction}

Magnetars are neutron stars with very high surface magnetic fields and frequent starquakes \citep{Duncan1998}. Unlike rotation-powered neutron stars, magnetars are powered by their strong magnetic fields, instead of their spin-down energy~\citep{Duncan1992}. The typical magnetic field of magnetars is $>10^{14}$~G; however, the discovery of a low-magnetic-field SGR 0418+5729 put the lower limit  to be $7.5\times10^{12}$ G \citep{Rea2010}. 

During the past decade, more than 100 \gr~pulsars (mostly young pulsars and millisecond pulsars) have been identified through their \gr~pulsation, thanks to the unprecedented sensitivity at the 100~MeV to $>$300~GeV energy range of the \emph{Fermi} Large Area Telescope (LAT) \citep{Abdo2013} and multiwavelength observations. On the other hand, soft gamma-ray repeaters (SGRs) and anomalous X-ray pulsars (AXPs), both thought to be manifestations of magnetars, have not been seen at energies above several hundred keV. 

Most of the known $>$20 SGRs and AXPs~\footnote{McGill magnetar catalog: \url{http://www.physics.mcgill.ca/\~pulsar/magnetar/main.html}} are located at low Galactic latitudes \citep[][]{Olausen2014}, where the $\gamma-$ray contamination by the strong Galactic diffuse emission is severe. Also, in many cases, the presence of known supernova remnants (SNRs), molecular clouds, and/or energetic pulsars in the magnetars' neighborhood impose source confusion in hard $\gamma-$ray bands. Meanwhile, a series of GeV-bright SNR$-$molecular cloud (MC) association systems have been discovered with LAT \citep[e.g.,][]{Abdo2009, Abdo2010a, Abdo2010b, Abdo2010c, Castro2013, Xing2014, Liu2015, Araya2015}. Therefore, whether magnetars are intrinsically dark in MeV--GeV energies, or their GeV pulsations are just buried under the \gr~backgrounds, are still unclear. \citet{Abdo2010_magnetar_sample} analysed the first $\sim$17~months of LAT data of 13 magnetars and did not find convincing evidence for \grs~from any of the magnetar.

SGR 1806-20 was first discovered to be a source of soft $\gamma-$ray bursts \citep{Laros1986} and its bursts were found to be recurrent \citep{Atteia1987, Laros1987}. It is also famous for its 2004 December 27 giant flare~\citep{Hurley2005}. The persistent X-ray counterpart of SGR 1806-20, AX~1805.7-2025, was discovered by \citet{Murakami1994} with ASCA. The X-ray pulsation with a period of 7.47 s was determined by \citet{Kouveliotou1998} and a spin-down rate of $\sim2.6\times10^{-3}$ s yr$^{-1}$ was  found. \citet{Woods2000} used a series of RXTE observation to investigate the spin evolution of SGR 1806-20 and found that SGR 1806-20 contains a significant timing noise.  The spin history was refined by many investigations \citep[e.g.,][]{Mereghetti2000, Woods2007}. The latest long-term (years) spin history was reported by \citet{Younes2015} with  a spin-down rate of  $\sim2.53 \times 10^{-2}$ s yr$^{-1}$, which is larger than the historical values measured in 1995.

SGR 1806-20 is a member of the cluster of giant massive stars Cl* 1806-20~\citep{Fuchs1999, Figer2005, Corbel2004}, which is located within the giant Galactic H II complex W31~\citep{Corbel2004}. \citet{Bibby2008} determined that  Cl* 1806-20 has a distance of $8.7^{+1.8}_{-1.5}$ kpc from us  (which is consistent with a lower limit of 9.4 kpc set by \citet{Svirski2011}). Among the members of this stellar cluster is a luminous blue variable (LBV) hypergiant LBV 1806-20, which generates tremendous wind powering the radio nebula G10.0-0.3 at its core \citep[][]{Gaensler2001, Corbel2004}. Cl* 1806-20 also hosts four Wolf-Rayet (WR) stars and four OB supergiants~\citep{Eikenberry2004,Figer2005}. In the cluster, each WR star generates relatively intense wind, with a mass-loss rate of $\sim$10$^{-5.4}$-10$^{-4.2}$ $M_\odot$ yr$^{-1}$ and a terminal velocity of $\sim(1.2$-$3.1)\times10^6$ m s$^{-1}$ \citep[cf. Table 4 of ][]{Nugis2002}. %Although LBV 1806-20 generates even stronger wind, the four WR-stars in the cluster should also be considered as potential candidates of cosmic ray origins.

%Both G10.0-0.3 and SGR 1806-20 are, respectively, strong candidates for contributing to the hadronic and leptonic $\gamma-$ray emission at 3FGL J1809.2-2016c. 

A radio nebula  in W31, G10.0-0.3 \citep{Kulkarni1994} which has a luminosity of 10$^{32}$~erg~s$^{-1}$ at the distance of 8.7~kpc \citep{Bibby2008}, is believed to be powered by LBV 1806-20 where the radio flux peaks~\citep{Gaensler2001,Kaplan2002}, while analyses of Cerro Tololo Inter-American Observatory (CTIO) infrared, \emph{Chandra} X-ray and Inter-Planetary Network (IPN) $\gamma-$ray data  for SGR 1806-20 confirmed the magnetar position to be offset by $\gtrsim12"$ from the center of G10.0-0.3 \citep{Hurley1999, Eikenberry2001, Kaplan2002}. Also, VLA observations of G10.0-0.3 showed \emph{no} evidence of a blast wave or a supernova explosion because of a centrally condensed, time-varying morphology and an extraordinarily steep spectrum \citep{Kulkarni1994, Vasisht1995, Frail1997}. Therefore, \citet{Gaensler2001} doubted the putative SNR nature of this radio nebula and suggested that \emph{no} known SNR is associated with SGR 1806-20. 

HESS J1808-204 detected at the TeV band has an extended feature similar in scale and orientation to that of  G10.0-0.3, and hence they are argued to be associated with each other~\citep{Rowell2012}. Its 0.5-5~TeV energy flux of 1.3$\times$10$^{-12}$~erg~cm$^{-2}$s$^{-1}$ can readily be explained by the intense stellar wind from LBV~1806-20 from an energetic point of view~\citep{Rowell2012}. At \emph{Fermi}/LAT energies, a `confused' source, 2FGL J1808.5-2037c~\citep{Nolan2012} is catalogued at the southern edge of HESS J1808-204~\citep[cf. Figure 1 of ][]{Rowell2012}, while an updated \emph{Fermi}/LAT catalog \citep[3FGL;][]{Acero2015a} shows a `confused' source, 3FGL J1809.2-2016c, to the north-east of Cl* 1806-20. This highlights the complexity of the MeV--GeV emission from this region, and a dedicated investigation using all available LAT data is crucial to identify the origin of high-energy \gr~emission. %The MeV-TeV spectral connection was suggested by a single power-law in this field of view \citep[cf. Figure 2 of ][]{Rowell2012}, and a hadronic radiation mechanism can be reasonable to describe its $\gamma-$ray emission.  Therefore, the origin of hadronic cosmic rays responsible for TeV emission in this region is still unclear.

SNR G9.7-0.0, which is a shell-type non-thermal SNR \citep[][]{Frail1994, Brogan2006}, is separated from Cl* 1806-20 by only $\sim0.35^\circ$ as projected on the sky. However, its distance from us of 4.7 kpc \citep[][]{Hewitt2009} is inconsistant with that of  Cl* 1806-20, making  it impossible for them to be related to each other. The MC interaction of this SNR has been confirmed by the detection of a nearby OH(1720 MHz) maser \citep[][]{Hewitt2009}, and hence it is a potential candidate for $\gamma-$ray emission.

In this work, we explore the MeV-GeV emission in the field of SGR~1806-20/Cl*~1806-20 by using $\sim$7 years of \emph{Fermi} LAT data with the latest instrumental responses and background models. Then, we compare its morphology and spectrum to those of HESS J1808-204 (which is associated with G10.0-0.3). We also  examine the correlation between the long-term temporal behavior of LAT flux and the X-ray outburst history of SGR 1806-20. In turn, we provide some insight into the possible origin(s) of the $\gamma-$rays.

\section{Observation \& Data Reduction}

We performed a series of binned maximum-likelihood analyses for a 20$^\circ$$\times$20$^\circ$ ROI centered at RA=$18^{h}08^{m}11.277^{s}$, Dec=$-20^\circ28'52.82"$ (J2000), which is the centroid of 1-50 GeV emission around 3FGL J1809.2-2016c. We  used the data obtained by LAT between 2008 August 4 and 2015 September 3. The data were reduced and analyzed with the aid of \emph{Fermi} Science Tools v10r0p5 package. In view of the complicated environment of the Galactic plane region, we adopted the events classified as Pass8 ``Clean" class for the analysis so as to better suppress the background. The corresponding instrument response function (IRF) ``P8R2$_-$CLEAN$_-$V6" is used throughout the investigation.

Considering that we  include  photons with energies 60-300 MeV, and that we are investigating a crowded region on our Galactic plane, we focused on the events belonging to either ``FRONT" or ``PSF3" partition for better spatial resolution. In those cases which favor spectral resolution and/or photon statistics more than spatial resolution, we adopted ``FRONT" data instead of ``PSF3" data. We further filtered the data by accepting only the good time intervals where the region-of-interest (ROI) was observed at a zenith angle less than 90$^\circ$ so as to reduce the contamination from the albedo of Earth.

For subtracting the background contribution, we have included the Galactic diffuse background (gll$_-$iem$_-$v06.fits), the isotropic background (iso$_-$P8R2$_-$CLEAN$_-$V6$_-$PSF3$_-$v06.txt for ``PSF3" data or iso$_-$P8R2$_-$CLEAN$_-$V6$_-$FRONT$_-$v06.txt for ``FRONT" data) as well as all other  point sources cataloged in 3FGL within 25$^\circ$ from the ROI center in the source model. We  set free the spectral parameters of the 3FGL sources within 7$^\circ$ from the ROI center in the analysis. For the  3FGL sources beyond 7$^\circ$ from the ROI center, their spectral parameters were fixed at the catalog values. 

In spectral and temporal analysis, we required each energy-bin and time-segment to attain a signal-to-noise ratio $>3\sigma$ for a robust result. For each energy-bin or time-segment \emph{dissatisfying} this requirement, we placed a 2$\sigma$ upper limit on its flux.

\section{Data Analysis}
\subsection{Spatial Analysis}

\begin{figure}
	\centering\vspace{-5.8cm}
	\includegraphics[height=1.1\linewidth]{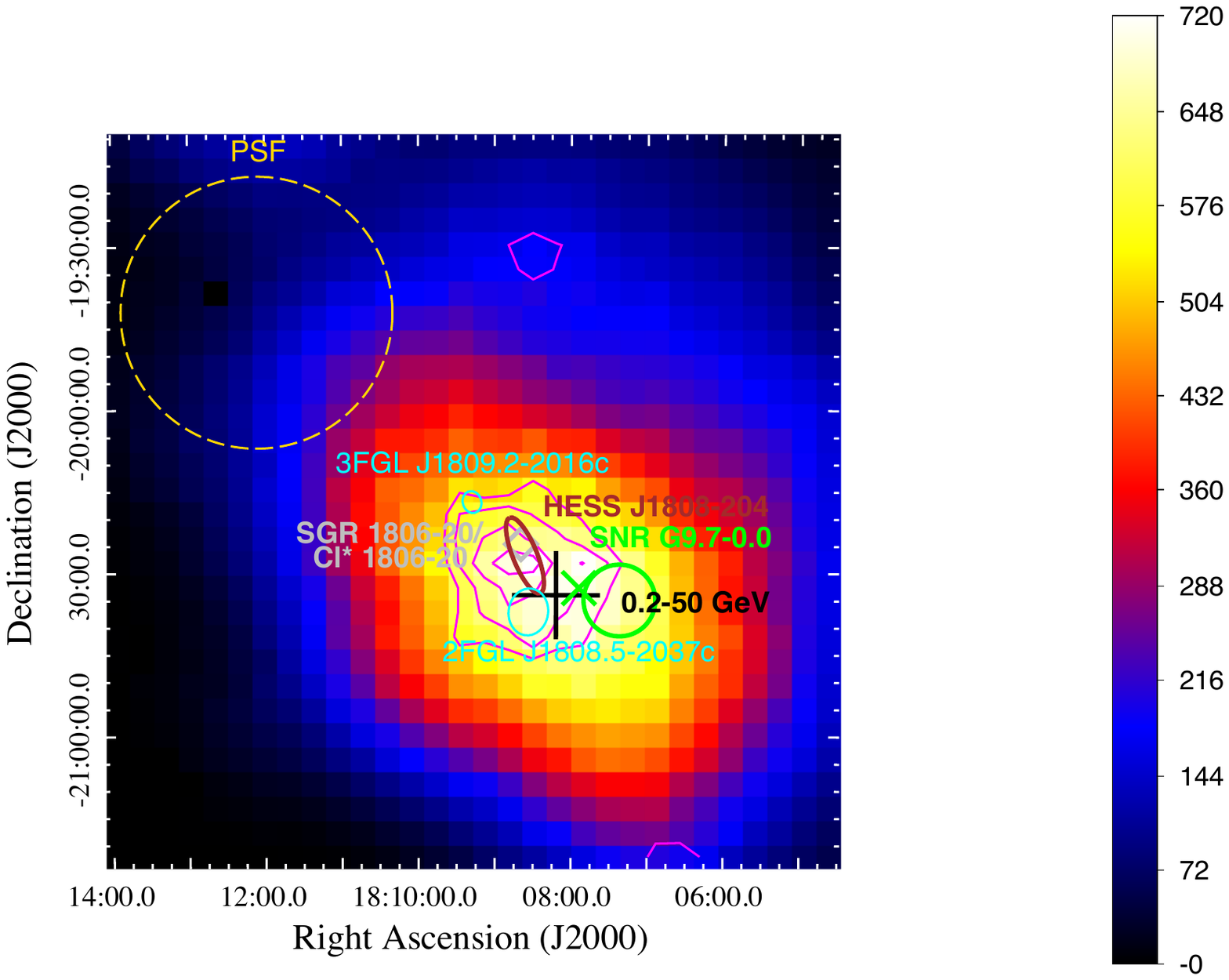}\vspace{-10cm}\\
	\includegraphics[height=1.1\linewidth]{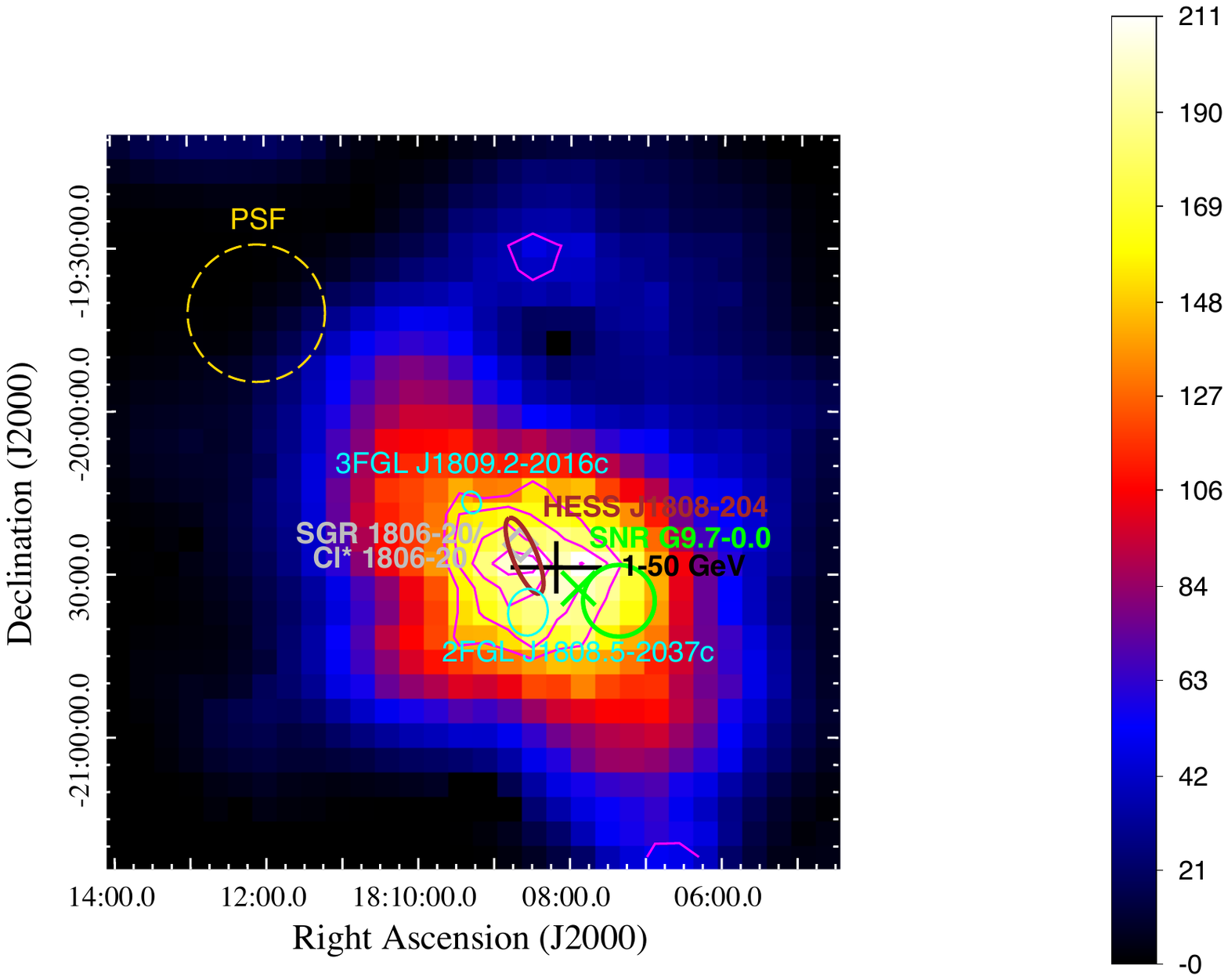}\vspace{-5.8cm}\\
	\caption{The TS maps in 0.2-50 GeV (top) and 1-50 GeV (bottom) respectively, where all neighbouring 3FGL catalog sources except 3FGL J1809.2-2016c are subtracted. On each map, the 95\% confidence region of the centroid is indicated as a black thick cross, and the FWHM of the PSF is illustrated by a golden dashed circle.  Both panels are overlaid with the magenta contours of detection significance (4, 4.5, 5, 5.5$\sigma$) determined in  2.5-500 GeV. The position and extents of HESS J1808-204, described as a brown thick ellipse, are taken from \citet{Rowell2012}. The position and dimension of SNR G9.7-0.0, described by a green thick circle, are taken from \citet{Brogan2006}, and the position of its OH maser, indicated as a green ``X", is taken from \citep[][]{Hewitt2009}. The position of SGR 1806-20/Cl* 1806-20, indicated as a grey diamond, is taken from \citet{Israel2005}.}
	\label{SGR1806-20_tsmap1}
\end{figure}

The test-statistic (TS) maps of the field around 3FGL J1809.2-2016c for ``PSF3" data are shown in Figure~\ref{SGR1806-20_tsmap1}, where all 3FGL catalog sources except 3FGL J1809.2-2016c are subtracted.  The morphologies in 0.2-50 GeV and  1-50 GeV are both ellipse-like, with a major axis of $\sim45^\circ$ anti-clockwise from the north. The peak detection significance is $\sim27\sigma$ in 0.2-50 GeV and $\sim15\sigma$ in 1-50 GeV. The 95\% confidence regions of centroids determined on these two maps overlap more than one-third of the area of each other. They also overlap more than one-third of the area of the extents of HESS J1808-204. The centroid at  1-50 GeV is positionally consistent with SGR 1806-20/Cl* 1806-20, and both 0.2-50 GeV and 1-50 GeV centroids are positionally consistent with SNR G9.7-0.0 as well as its maser. The 1-50 GeV centroid is taken to be the center of our ROI.

\begin{figure}
	\centering
	\includegraphics[width=0.78\linewidth]{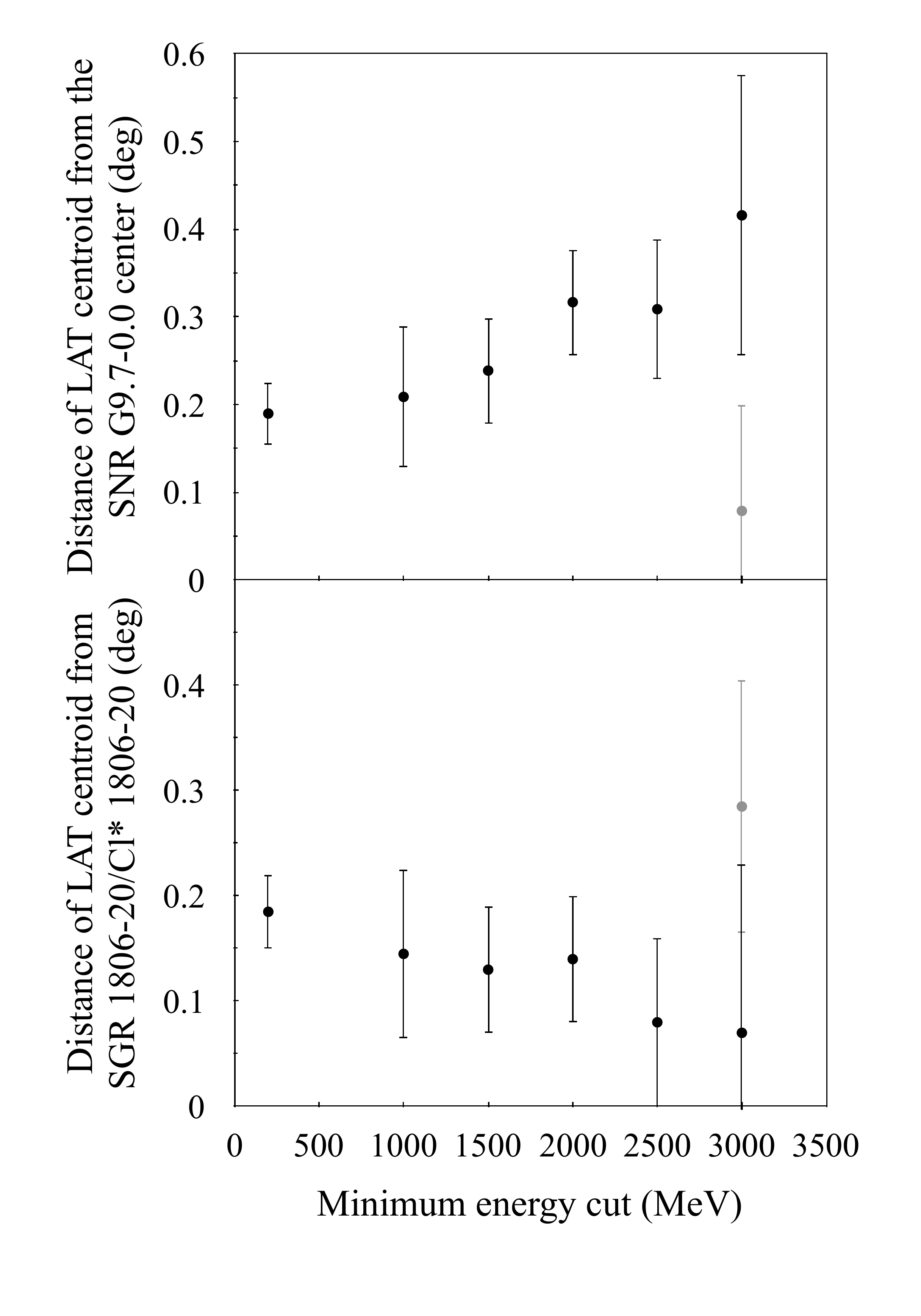}
	\caption{The centroid distances, measured from the SNR G9.7-0.0 center and SGR 1806-20/Cl* 1806-20 respectively, as  functions of the  $E_\mathrm{{cut,min}}$. At the $E_\mathrm{{cut,min}}$ of  3 GeV, the entire feature is resolved to be two separated clumps (with a $\sim3\sigma$ significance), so there are two data points  on each panel, where the gray one is for the fainter (less significant) clump. }
	\label{centroid}
\end{figure}

In order to examine  whether the centroid position is significantly dependent on the energy band, we also created TS maps with the minimum energy cut ($E_\mathrm{{cut,min}}$) shifted to 1.5, 2, 2.5 and 3 GeV, and the maximum energy cut  shifted to 500 GeV. The contours of detection significance (4, 4.5, 5, 5.5$\sigma$) determined on the 2.5-500 GeV TS map are overlaid on both panels of Figure~\ref{SGR1806-20_tsmap1}.  The centroid distances, measured from the SNR G9.7-0.0 center and SGR 1806-20/Cl* 1806-20 respectively, as  functions of the  $E_\mathrm{{cut,min}}$ are shown in Figure~\ref{centroid}.

The centroid at the $E_\mathrm{{cut,min}}$ of 200 MeV is almost equidistant ($\sim0.19^\circ$) from the SNR G9.7-0.0 center and SGR 1806-20/Cl* 1806-20. As the $E_\mathrm{{cut,min}}$ increases from 200 MeV to 2.5 GeV,  the distances of the centroid from the SNR G9.7-0.0 center and from SGR 1806-20/Cl* 1806-20 remain essentially constant ($\chi^2<5$ for 4 d.o.f.).  Nevertheless, one noticeable thing in 2.5-500 GeV is  that, the detection significance at HESS J1808-204 and SGR 1806-20/Cl* 1806-20 is $\gtrsim5.5\sigma$ while the detection significance at SNR G9.7-0.0 and its OH maser is $\lesssim4.5\sigma$. 

With the $E_\mathrm{{cut,min}}$ further pushed to 3 GeV, the entire feature  appears to be resolved into two separated clumps, each of which has  a   significant detection ($3.2-3.4\sigma$). Although the `dip' between their  centroids  is \emph{not} statistically significant ($<2.5\sigma$), it is noticeable that the regions of these two clumps are \emph{respectively} coincident with HESS J1808-204 and SNR G9.7-0.0.  In order to quantify the significance of two emission sites resolved in this energy band, we performed two tests: We re-made the TS map with the brighter clump modelled as an additional point source and subtracted, and we found that the residual at the other clump still has a detection significance of $\sim3.0\sigma$; in a likelihood ratio test, we found that a model with two point sources (representing the two clumps respectively) is preferred over that with a single point source (representing the brighter clump) by $\sim3.0\sigma$. Therefore, we  have  strong evidence for the two-emission-site morphology at energies $\gtrsim3$ GeV.

\begin{figure}
	\centering
	\includegraphics[width=.49\linewidth]{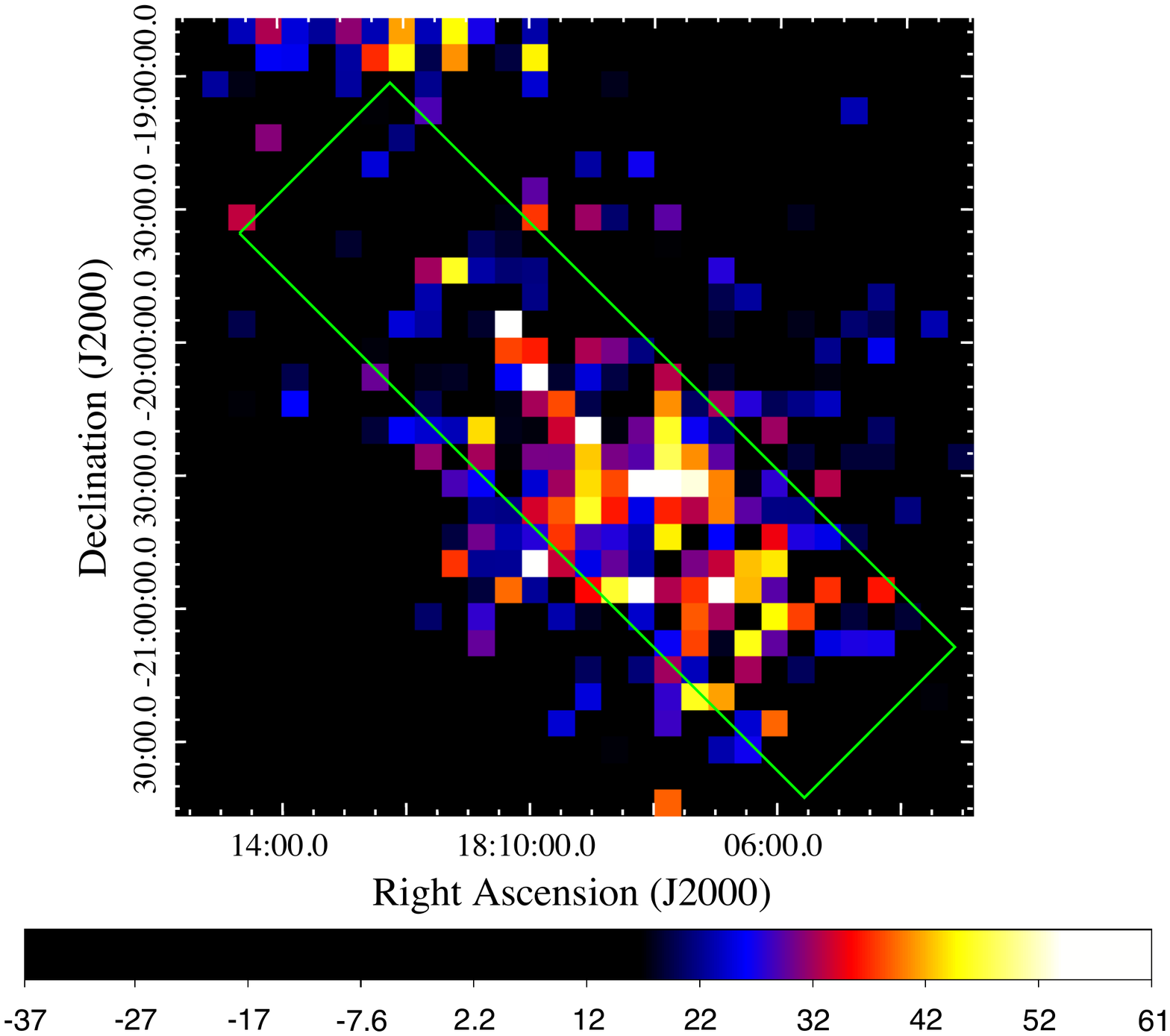}
	\includegraphics[width=.49\linewidth]{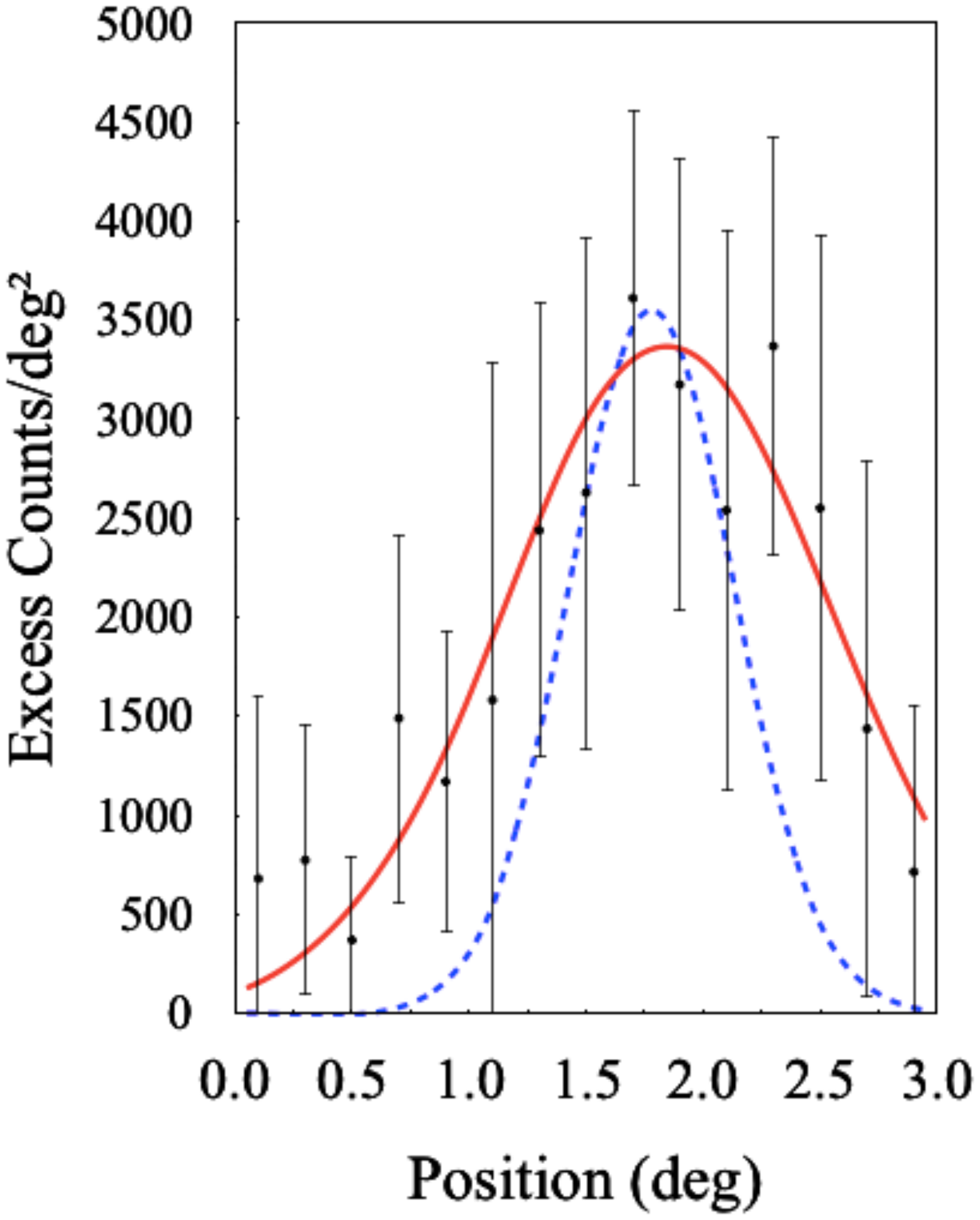}
	\caption{\textbf{Left:} A background-subtracted $\gamma-$ray count-map at 0.2-50 GeV, where all sources except 3FGL J1809.2-2016c are subtracted. It is superimposed with the chosen slice (in green) along the orientation of extension. \textbf{Right:} The plot of excess counts along the slice with the best-fit Gaussian (in red), overlaid with the simulated profile of an expected point source (the blue dashed Gaussian).}
	\label{SGR1806-20_bgsub}
\end{figure}

For further investigating the 0.2-50 GeV morphology of 3FGL J1809.2-2016c, we followed the scheme adopted by \citet{Hui2016}. We produced a  $\gamma-$ray count-map where all 3FGL catalog sources except 3FGL J1809.2-2016c are subtracted, and then computed a brightness profile along the major axis of the ellipse-like feature. We also simulated an expected point-like source with the same spectrum as 3FGL J1809.2-2016c. The result is shown in Figure~\ref{SGR1806-20_bgsub}. To examine the source extension, we have fitted the profile with a single Gaussian. It yields a FWHM of $1.65^\circ \pm 0.22^\circ$ ($\chi^2$ = 2.51 for 12 d.o.f.),  exceeding that of the simulated point source, $0.83^\circ$, by $>3.5\sigma$. %\footnote{The typical FWHM of a single point source at 540 MeV is $\sim1.09^\circ$, which is calculated from the corresponding 68\% containment radius provided by \url{http://www.slac.stanford.edu/exp/glast/groups/canda/lat$_-$Performance.htm}.}. 
We repeated this exercise for the minor axis of the ellipse-like feature, and obtained a FWHM of $1.53^\circ \pm 0.19^\circ$ ($\chi^2$ = 3.40 for 12 d.o.f.). This also exceeds that of the simulated point source by $>3.5\sigma$. These suggest that the MeV-GeV emission from 3FGL J1809.2-2016c is extended along both major and minor axes.

Since  the feature around 3FGL J1809.2-2016c is extended with the major and minor axes consistent within the tolerance of statistical uncertainties, we  replaced the `confused' point source 3FGL J1809.2-2016c with a  circularly extended source in the source model for subsequent analyses.  We named it \emph{Fermi} J1808.2-2029, assigned it a single power-law, and we attempted  uniform disks of different radii. They are centered at the 1-50 GeV centroid (our ROI center), which is determined with better spatial resolution and sufficient photon statistics. The values of the \emph{ln(likelihood)} in 0.2-50 GeV for ``FRONT" data are tabulated in Table~\ref{morphology}. We determined the radius to be $0^\circ.65^{+0^\circ.05}_{-0^\circ.04}$ and this morphology is preferred over a point-source model by $>15\sigma$. Therefore, we modelled \emph{Fermi} J1808.2-2029 as  a   uniform disk with   $0.65^\circ$ radius, in subsequent analyses.

\subsection{Spectral Analysis}

To construct the binned spectrum of \emph{Fermi} J1808.2-2029, we performed an independent fitting of each spectral bin adopting ``FRONT" data.  We examined how well the 0.2-50 GeV spectrum can be described by, respectively, a simple power-law (PL)
\begin{center}
$\frac{dN}{dE}=N_0(\frac{E}{E_0})^{-\Gamma}$ \ \ \ ,
\end{center}
an exponential cutoff power law  (PLE)
\begin{center}
$\frac{dN}{dE}=N_0(\frac{E}{E_0})^{-\Gamma}\mbox{exp}(-\frac{E}{E_\mathrm{c}})$ \ \ \ ,
\end{center}
and a broken power law (BKPL)
\begin{center}
	$\frac{dN}{dE}=\begin{cases} N_0(\frac{E}{E_\mathrm{b}})^{-\Gamma_1} & \mbox{if } E<E_\mathrm{b} \\ N_0(\frac{E}{E_\mathrm{b}})^{-\Gamma_2} & \mbox{otherwise} \end{cases}$ \ \ \ .
\end{center}

\begin{figure}
	\centering
	\includegraphics[width=0.98\linewidth]{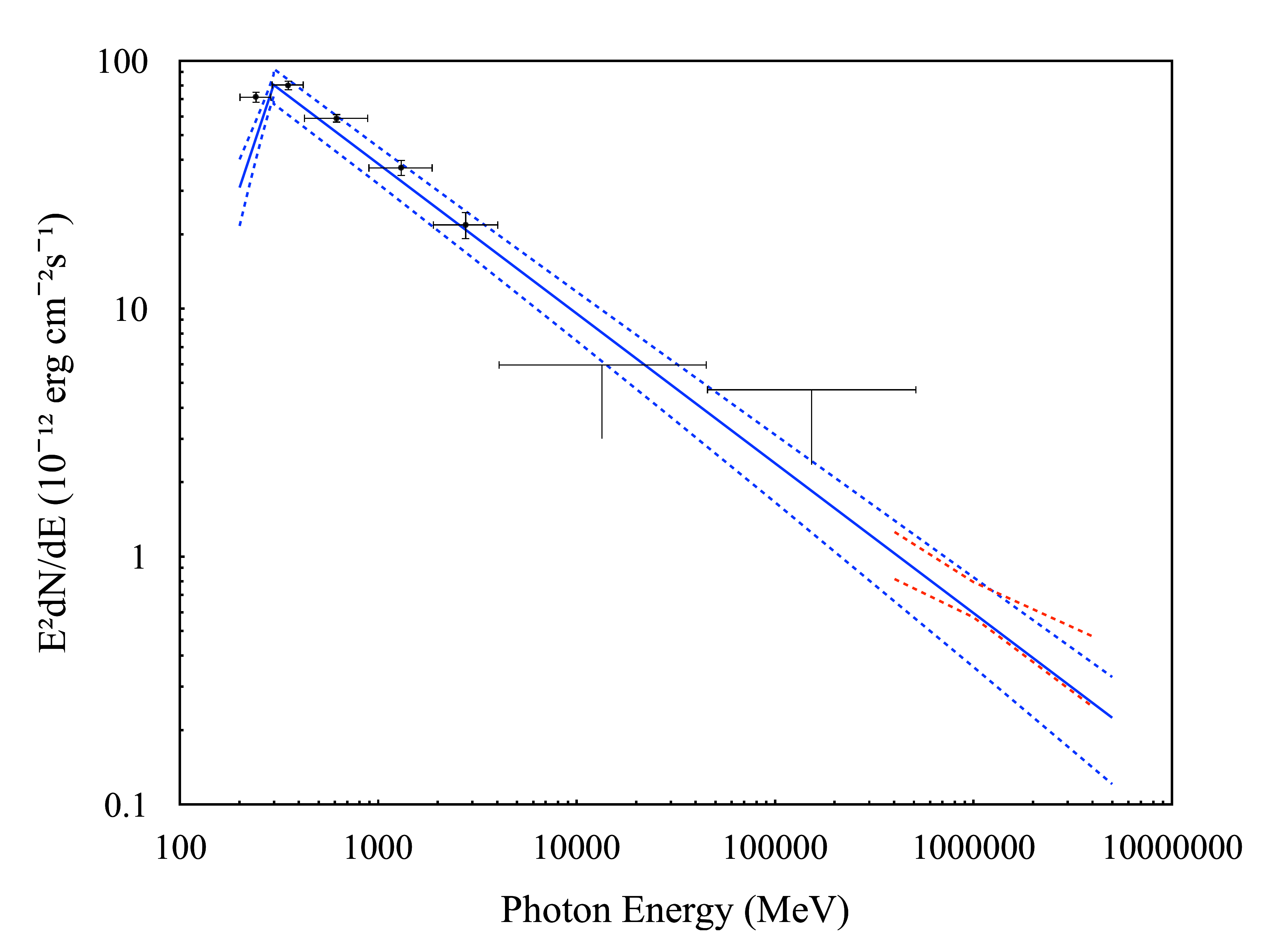}
	\caption{The SED of \emph{Fermi} J1808.2-2029/HESS J1808-204. The upper limits are at the $2\sigma$ confidence level. The  blue  lines  illustrate the best-fit  BKPL model (solid) as well as the range allowed by a 1$\sigma$ uncertainty (dashed) in 0.2-50 GeV.  Sandwiched between the two red dashed curves is the 0.4-4 TeV HESS flux allowed by a 1$\sigma$ uncertainty \citep[cf. Figure 2 of ][]{Rowell2012}.}
	\label{SGR1806-20_SED}
\end{figure}

For each spectral bin, we assigned \emph{Fermi} J1808.2-2029 a PL model. The results of spectral fitting are tabulated in  Table~\ref{spectral}, and the spectral energy distribution (SED)  is shown in Figure~\ref{SGR1806-20_SED}.

In 0.2-50 GeV, the likelihood ratio test indicates that PLE is preferred over PL by $\sim6.5\sigma$. A PLE model yields a photon index of $\Gamma=2.09 \pm 0.08$ and a cutoff energy of $E_\mathrm{c}=3628 \pm 1017$ MeV.  BKPL is preferred over PL by $\sim8.0\sigma$, and the TS value BKPL yields is higher than that  PLE yields by $\sim26$. Despite the poorly constrained  index $\Gamma_{1}=-0.41 \pm 0.71$ below the spectral break, the spectral break and the index above the break are well constrained to be $E_\mathrm{b}=297 \pm 15$ MeV and $\Gamma_{2}=2.60 \pm 0.04$. The spectrum  above $E_\mathrm{b}$ is steeper than that below $E_\mathrm{b}$ by  $>4 \sigma$.

Extrapolating the BKPL model to 0.4-4 TeV, we obtain  an estimated flux consistent with the H.E.S.S. measurements \citep[reported by][]{Rowell2012}, within the tolerance of statistical uncertainties (see Figure~\ref{SGR1806-20_SED}). 

\subsection{Temporal Analysis}

%\subsubsection{Long-term Variability}

\begin{figure}
	\centering
	\includegraphics[width=.98\linewidth]{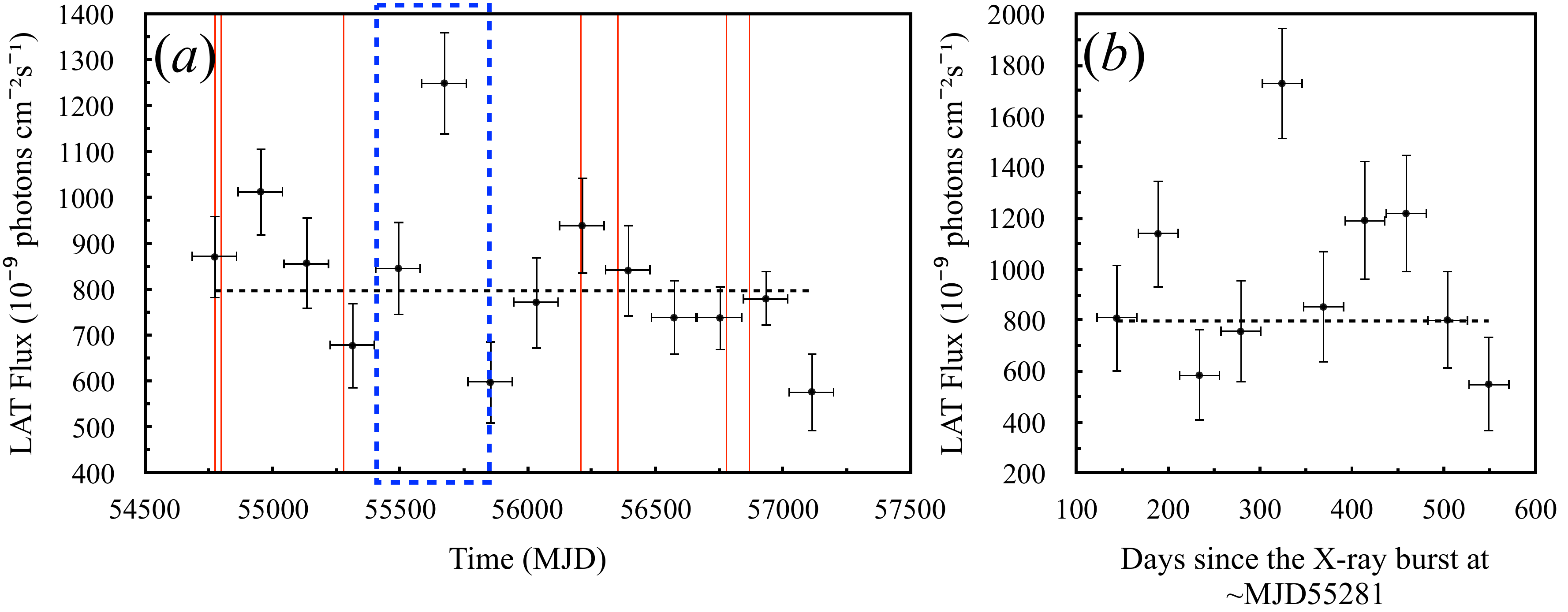}
	\caption{($a$) The light-curve of \emph{Fermi} J1808.2-2029 with 180 days as a segment.  The dashed horizontal line indicates the $\sim$6.9-year average flux (the best-fit uniform distribution to this light-curve). The red vertical lines indicate the dates of X-ray outbursts of SGR 1806-20. The time range within the blue dashed box was further divided into 10 45-day segments. ($b$) The light-curve of \emph{Fermi} J1808.2-2029 with 45 days as a segment.  The $\sim$6.9-year average flux is also indicated by a dashed horizontal line.}
	\label{SGR1806-20_lc}
\end{figure}

In order to examine the long-term variability of \emph{Fermi} J1808.2-2029, we divided the first $\sim$6.9 years of \emph{Fermi} LAT observation into 14 180-day segments. A binned maximum-likelihood analysis of ``PSF3" data in 60 MeV - 50 GeV was performed for each individual segment. We assumed a PL model for \emph{Fermi} J1808.2-2029. %Considering the large systematic uncertainties of spectral shapes in using an individual PSF below 100 MeV, we are \emph{not} interested in the temporal behavior of the spectral shape of \emph{Fermi} J1808.2-2029. 
The temporal behavior of the photon flux  of \emph{Fermi} J1808.2-2029 is plotted with the X-ray outburst history of SGR 1806-20, taken from  GCN Circulars~\footnote{\url{http://gcn.gsfc.nasa.gov/gcn3\_archive.html}} and \citet{Collazzi2015}, altogether in Figure~\ref{SGR1806-20_lc}($a$). 

The  $\chi^2$ test indicates that the photon flux deviates from a uniform distribution at a confidence level of  $\sim99.98\%$ ($\chi^2$ = 39.03 for 13 d.o.f.), but the temporal variability shows \emph{no} correlation with the X-ray outburst history of SGR 1806-20.  Noticeably, the photon flux from MJD55582.655 to MJD55762.655 (in the $\sim302-482$ days after the X-ray outburst at $\sim$MJD55281) is greater than the $\sim$6.9-year average (the best-fit horizontal line) by  $\sim4.0$ times its statistical error.  If we randomly generate 14 data points of a Gaussian probability distribution with a mean and standard deviation based on the observed light-curve, in each of $10^6$ Monte-Carlo simulations, the chance probability to obtain at least one data point different from the average by $>4$ times its statistical error is $<0.1\%$.  This might indicate that our detection of the flux increment is \emph{not} an occasional chance event. 

In order to examine the gradualness or abruptness of such a flux increment, we divided the data $\sim122-572$ days after that X-ray outburst into 10 45-day segments, and performed a binned maximum-likelihood analysis for each segment. The temporal behavior of the photon flux  is shown in Figure~\ref{SGR1806-20_lc}($b$). The photon flux from MJD55582.655 to MJD55627.655 (in the data $\sim302-347$ days after that X-ray outburst)  is higher than the $\sim$6.9-year average by  $\sim4.3$ times its statistical error. Since the photon flux  within these 45 days is even higher than those in the $\sim302-482$ days after that X-ray outburst by  $\sim2.0$ times the statistical error, the flux increment is more likely to be abrupt. 

%We further divided the 287-362 days after that X-ray outburst into 5 15-day segments, and performed a binned maximum-likelihood analysis for each segment. As shown in Figure~\ref{SGR1806-20_lc}($c$), the photon flux is persistently high in the 302-347 days after the outburst, but in a low state in the two other segments. Therefore, we constrain the timescale of the flux increment to be $\lesssim$45 days.

In order to quantify the change of the photon flux and the spectral shape in the data $\sim302-347$ days after that X-ray outburst, we repeated the binned maximum-likelihood analysis in these 45 days with ``FRONT" data of energies 200 MeV - 50 GeV. As a result, a PL yields a signal-to-noise ratio of $\sim6.0\sigma$, which is sufficiently high for us to claim a significant detection, with a photon index of $\Gamma =2.72 \pm 0.24$ and a photon flux of $(2.20 \pm 0.39) \times 10^{-7}$ photons cm$^{-2}$ s$^{-1}$. The additional spectral parameters in PLE/BKPL are not statistically required based on a likelihood ratio test ($< 1\sigma$). Compared to the $\sim$7-year average values of BKPL parameters shown in the Table~\ref{spectral}, the 200-300 MeV spectral shape becomes steeper at a $>4\sigma$ level, the 0.3-50 GeV photon index is consistent with the $\sim$7-year average within the tolerance of statistical uncertainties, and the flux rises by  $\sim1.6$ times the statistical error. In order to further check the robustness, we repeated the aforementioned analysis with the spectral parameters of the Galactic diffuse background and isotropic background fixed at the $\sim$7-year averages. As a result, the photon index and photon flux both altered by only $\lesssim5\%$. 

We confirm a genuine LAT flux enhancement of \emph{Fermi} J1808.2-2029 within the 45 days. Since  the ratio of the flux increment to the statistical error drops from $\sim4.3$  at $>$60 MeV to $\sim1.6$  at $>$200 MeV and the 200-300 MeV spectral shape becomes much steeper in these  45 days,  we infer that almost the entire enhancement occurs  at energy $<$400 MeV. %Furthermore, the 95\% confidence region  in these 45 days, determined from  ``PSF3" data of energy within 0.2-50 GeV, is consistent with the 1-50 GeV centroid in the $\sim$7 years.

%\subsubsection{Pulsation search}

%We performed pulsation search for SGR 1806-20 using the all-partition data  collected from 2011 January 21 to 2011 Mar 7 (the 45 days of the dramatic LAT flux enhancement). We searched in two energy ranges, 20-400 MeV and 20-800 MeV, respectively.

%The analyses of \emph{XMM-Newton} observations determined the spin-frequency of SGR 1806-20 on 2011 Mar 23 to be $\nu=0.129838$ Hz, and the average spin-down rate from 2005 July to 2011 March to be $\dot{\nu}=1.35\times10^{-11}$ Hz s$^{-1}$ \citep[][]{Younes2015}. We adopted these $\nu$ and $\dot{\nu}$ as ephemeris information. The position of SGR 1806-20  reported by \citet{Israel2005} was taken for barycentric correction.

\section{Discussion}

% Tam->Paul: the following paragraph is too general, it can be put into the discussion/interpretation section.
\emph{Fermi} J1808.2-2029 is  extended mainly towards the south-west direction  (cf. Figures~\ref{SGR1806-20_tsmap1} \&~\ref{SGR1806-20_bgsub}).  %The mutually similar orientations of major extensions of \emph{Fermi} J1808.2-2029,  HESS J1808-204 and the radio nebula G10.0-0.3 \citep[cf. Figure 1 of ][]{Rowell2012} indicate that a significant portion of $\gamma-$rays may be associated with Cl* 1806-20.
Its emission region partly coincides with SNR G9.7-0.0 and partly coincides with SGR 1806-20/Cl* 1806-20. The association between Fermi J1808.2-2029 and HESS J1808-204,  where the latter  has been associated with the radio nebula G10.0-0.3, is  suggested by the  connection of 300 MeV - 4 TeV spectrum by a PL (cf. Figure~\ref{SGR1806-20_SED}). 

Leptonic particles can be accelerated in the outer gap (outer magnetosphere) region and/or pulsar wind region of a neutron star, and can then produce $\gamma-$rays through the curvature radiation process and inverse-Compton scattering (IC) of soft photons \citep[][]{Lyutikov2012,  Harding2015, Aharonian2012}. Hadronic particles are mostly accelerated by SNR shocks, and then collide with  protons in MCs to produce $\gamma-$rays through neutral-pion-decay \citep[][]{Ackermann2013}. Leptonic cosmic-rays generally emit $\gamma-$rays at lower energies than hadronic cosmic-rays due to synchrotron cooling of leptons. 

There are abundant infrared and optical photons within Cl* 1806-20 \citep[cf.][]{Kosugi2005, Balman2003, Israel2005, Rea2005}, and cosmic microwave background photons are everywhere. These both can be seed photons for leptonic cosmic-rays to produce $\gamma-$ray emission through IC. NANTEN survey reveals some CO  clouds positionally consistent with the 1-50 GeV centroid and/or 0.2-50 GeV centroid of \emph{Fermi} J1808.2-2029 \citep[cf. Figures 4($f$) and 17 of][]{Takeuchi2010}. These clouds can be collision sites for hadronic cosmic-rays to produce $\gamma-$ray emission. During the searching process for the magnetar SGR 1806-20 in the MeV-GeV band, distinguishing between $\gamma-$rays produced by leptonic and hadronic cosmic-rays is an important issue.

\subsection{Hadronic Scenario}

\subsubsection{Relations with SNR G9.7-0.0}

The spectrum of \emph{Fermi} J1808.2-2029 has a turnover at energies below 1 GeV, consistently with the \emph{Fermi} LAT spectra of  shell-type SNRs interacting with MCs such as W51C, W44, IC 443 and W28 (cf. Figure 3 of \citet{Abdo2009}; Figure 3 of \citet{Abdo2010b}; Figure 3 of \citet{Abdo2010c}; Figure 3(a) of \citet{Abdo2010a}). Therefore, significant $\gamma-$ray contribution from shell-type SNR G9.7-0.0 is suggested.

In 0.2-50 GeV, the most preferable spectral model for \emph{Fermi} J1808.2-2029, BKPL, yields a spectral index $\Gamma_{2}$ well within the range for GeV sources of  SNR$-$MC hadronic interaction \citep[cf.  Table 3 of ][]{Liu2015}. Integrations  adopting the BKPL parameters in Table~\ref{spectral} give $\gamma-$ray energy fluxes of $F(>E_\mathrm{b})\sim2.19\times10^{-10}$ erg cm$^{-2}$ s$^{-1}$ and $F($1-100 GeV$)\sim9.88\times10^{-11}$ erg cm$^{-2}$ s$^{-1}$. Assuming that \emph{Fermi} J1808.2-2029 is just next to SNR G9.7-0.0 (at a distance of $\sim4.7$ kpc from us), we obtain $\gamma-$ray luminosities of $L(>E_\mathrm{b})\sim5.80\times10^{35}$ erg s$^{-1}$ and $L($1-100 GeV$)\sim2.61\times10^{35}$ erg s$^{-1}$. Both the $L($1-100 GeV$)$  and  $L(>E_\mathrm{b})$ are  well within the ranges of  luminosities for  SNRs, according to   Table 3 of \citet{Liu2015} and \citet{Bamba2015} respectively.

However, the $\gamma-$ray spectra of many GeV-detected SNRs have a spectral break at a few GeV \citep[][]{Acero2015b}, in  constrast to the PL connection of 300 MeV - 4 TeV spectrum of \emph{Fermi} J1808.2-2029/HESS J1808-204.  Noticeably, the 2.5-500 GeV detection significance at SNR G9.7-0.0 and its OH maser drops to $\lesssim4.5\sigma$ (cf. Figure~\ref{SGR1806-20_tsmap1}), and the region of HESS J1808-204 is totally inconsistent with that of SNR G9.7-0.0. Therefore, the interacting supernova remnant SNR G9.7-0.0 can only account for the $\gamma-$ray emission from 200 MeV to several GeV, but is \emph{unlikely} to contribute significantly to the emission at energies above several GeV.

\subsubsection{Relations with Cl* 1806-20}

 There are a number of MCs along the line of sight towards \emph{Fermi} J1808.2-2029 and Cl* 1806-20, including MC 73 and MC -16, whose distances are consistent with that of Cl* 1806-20, i.e., $\sim8.7$ kpc \citep[][]{Corbel2004, Takeuchi2010}. They can be collision sites for hadronic cosmic-rays from Cl* 1806-20 to produce $\gamma-$ray emission.

Assuming that \emph{Fermi} J1808.2-2029 is just next to Cl* 1806-20 (at a distance of $\sim8.7$ kpc from us), we obtain $\gamma-$ray luminosities of $L(>E_\mathrm{b})\sim1.99\times10^{36}$ erg s$^{-1}$ and $L($1-100 GeV$)\sim8.95\times10^{35}$ erg s$^{-1}$. The $L($1-100 GeV$)$ is marginally within the range for GeV sources of  SNR$-$MC hadronic interaction, while the $L(>E_\mathrm{b})$ is beyond the range of 0.1-100 GeV luminosities for  SNRs. We assume the average number density of protons in MCs near Cl* 1806-20 to be 100 cm$^{-3}$, which is appropriate for MC 73 and MC -16 \citep[cf.][]{Corbel2004}. The  cross section area of proton-proton collisions is $\sim10^{-26}$ cm$^2$, and the angular diameter  of the cloud is $\sim0.15^\circ$,  which corresponds to $\sim$20~pc at 8.7kpc. We also assume the $\gamma-$ray conversion efficiency for each individual proton-proton collision to reach the maximum of 0.1. Hence, we inferred the $\gamma-$ray conversion efficiency of cosmic-ray energy  to be $\sim7.0\times10^{-6}$ and the required power  from a nearby cosmic-ray accelerator to be $P^\mathrm{local}_\mathrm{CR}\sim2.8\times10^{41}$ erg s$^{-1}$. 

A typical supernova explosion releases energy of a canonical amount of $\sim10^{51}$ erg, and its remnant can vigorously accelerate cosmic rays for $>5$ kyr \citep[][]{Dermer2013}, with an efficiency of $\sim10\%$ for converting kinetic energy to non-thermal cosmic-ray energy \citep[][]{Ginzburg1964}. Therefore, the energy budget $P^\mathrm{local}_\mathrm{CR}\sim2.8\times10^{41}$ erg s$^{-1}$ is  so high that even a combined contribution from  several SNRs inside or around Cl* 1806-20, if  they exist,  \emph{cannot} supply it.

 Even if SGR 1806-20 is a GeV-emitting magnetar, it normally accelerates leptons but not hadrons, like other $\gamma-$ray pulsars \citep[cf.][]{Abdo2013}. Therefore, it is reasonable to exclude  SGR 1806-20 as a major hadronic source of \emph{Fermi} J1808.2-2029. 
%The analyses of \emph{XMM-Newton} observations determined the spin-frequency of SGR 1806-20 on 2011 Mar 23 to be $\nu=0.129838$ Hz, and the average spin-down rate from 2005 July to 2011 March to be $\dot{\nu}=1.35\times10^{-11}$ Hz s$^{-1}$ \citep[][]{Younes2015}. Hence, we obtain a spin-down power of $L_\mathrm{sd}\sim6.92\times10^{34}$ erg s$^{-1}$.  Assuming that SGR 1806-20 is the counterpart of \emph{Fermi} J1808.2-2029, we obtain an  efficiency of cosmic-ray production of $P^\mathrm{local}_\mathrm{CR}/L_\mathrm{sd}\sim4.1\times10^6$. Therefore, it is reasonable to exclude the spin-down of SGR 1806-20 as the major source of \emph{Fermi} J1808.2-2029. 
%
%Adopting $\nu=0.129838$ Hz and $\dot{\nu}=1.35\times10^{-11}$ Hz s$^{-1}$, we  also obtain a surface magnetic field strength of $B\sim5.03\times10^{15}$ G at the pole. Hence, we can estimate the power of magnetic field decay to be $L_\mathrm{B}>10^{36}$ erg s$^{-1}$ \citep[cf. ][]{Zhang2003}. Assuming that SGR 1806-20 is the counterpart of \emph{Fermi} J1808.2-2029, we obtain an  efficiency of cosmic-ray production of $P^\mathrm{local}_\mathrm{CR}/L_\mathrm{B}<2.8\times10^5$. Therefore, the loss of magnetic energy of SGR 1806-20 is also \emph{impossible} to account for the emission detected at \emph{Fermi} J1808.2-2029.

\citet{Rowell2012} constrained  the total kinetic energy of all stellar winds  from Cl* 1806-20 to be $L_\mathrm{w}>10^{38}$ erg s$^{-1}$, which is dominated by LBV 1806-20 and/or the four WR-stars. Assuming that the entire cluster is the  energy source of \emph{Fermi} J1808.2-2029, we obtain an  efficiency of cosmic-ray production of $P^\mathrm{local}_\mathrm{CR}/L_\mathrm{w}<3000$. Therefore,  there is  \emph{no} evidence for the combined stellar wind of all Cl* 1806-20 members  to be the major source. 

Regardless of the cosmic-ray origin(s), the proton density ($\sim$100 cm$^{-3}$) in MCs near Cl* 1806-20 is far from being sufficient to cause the observed $\gamma-$ray emission. It follows that a purely hadronic scenario does \emph{not} support the adjacence between \emph{Fermi} J1808.2-2029  and this cluster at all.

\subsection{Leptonic Scenario}

%Another preferable spectral model within 0.2-50 GeV, PLE, yields a spectral index marginally within the range of typical values of about 1-2 in the 2nd pulsar catalog, and a cutoff energy  within the range of typical values of about 1-4 GeV \citep[][]{Abdo2013}. Therefore, we \emph{cannot} totally reject the leptonic contribution from the magnetar SGR 1806-20 either.

 The analyses of \emph{XMM-Newton} observations determined the spin-frequency of SGR 1806-20 on 2011 Mar 23 to be $\nu=0.129838$ Hz, and the average spin-down rate from 2005 July to 2011 March to be $\dot{\nu}=1.35\times10^{-11}$ Hz s$^{-1}$ \citep[][]{Younes2015}. Hence, we obtain a spin-down power of $L_\mathrm{sd}\sim6.92\times10^{34}$ erg s$^{-1}$. 
Adopting the same $\nu$  and $\dot{\nu}$, we  also obtain a surface magnetic field strength of $B\sim5.03\times10^{15}$ G at the pole. Hence, we can estimate the power of magnetic field decay to be $L_\mathrm{B}>10^{36}$ erg s$^{-1}$ \citep[cf. ][]{Zhang2003}. Here, we have $L_\mathrm{B}>10L_\mathrm{sd}$, which is consistent with the prediction for magnetars by \citet{Duncan1992}. Assuming that SGR 1806-20 is the energy source of \emph{Fermi} J1808.2-2029, we obtain  $\gamma-$ray conversion efficiencies of   $L(>E_\mathrm{b})/L_\mathrm{sd}\sim29$ and  $L(>E_\mathrm{b})/L_\mathrm{B}<2.0$. 
Assuming that the entire cluster is the energy source of \emph{Fermi} J1808.2-2029, we obtain a $\gamma-$ray conversion efficiency of $L(>E_\mathrm{b})/L_\mathrm{w}<0.02$. 
Therefore, the total kinetic energy of all stellar winds from Cl* 1806-20 can easily account for the emission detected at \emph{Fermi} J1808.2-2029,  while  the energy loss of SGR 1806-20 alone,  which mostly arises from magnetic energy,  can only contribute to a small component of the emission. 

Whereas, synchrotron cooling generally makes leptonic cosmic-rays difficult to produce $\gamma-$ray photons of a few GeV or above  via synchrotron radiation. It follows that normal stellar winds from Cl* 1806-20 \emph{cannot} explain the strong emission at energies above a few GeV and the GeV-TeV PL connection. However, with the reduced synchrotron losses for high-energy IC-emitting electrons, pulsar wind nebulae (PWNe) can maintain their high GeV-TeV $\gamma-$ray fluxes for timescales exceeding the lifetime of their progenitor pulsars \citep[][]{Tibolla2011}. Most TeV-detected PWNe are associated with pulsars of high spin-down power $>10^{36}$ erg s$^{-1}$ \citep[][]{Halpern2010}. Although the spin-down power of SGR 1806-20 is an order of magnitude lower than this threshold, the major mechanism of energy injection for a magnetar is the rapid decay of its strong magnetic field \citep[][]{Duncan1992}, which may account for the PWN-required power for SGR 1806-20. 

Assuming that the loss of magnetic energy of SGR 1806-20 is the major source for the emission at energies $>4$ GeV, an integration  adopting the BKPL parameters in Table~\ref{spectral} yields a $\gamma-$ray conversion efficiency of $L(>4$ GeV$)/L_\mathrm{B}<0.41$. Therefore, SGR 1806-20 alone is sufficient to generate a PWN which may account for the flux at energies $>4$ GeV.  Furthermore, the GeV-TeV spectral connection %, and the similar orientations of major extensions of \emph{Fermi} J1808.2-2029 and HESS J1808-204, are
is also consistent  with this PWN scenario. Noticeably, the photon index $2.39 \pm 0.19$ of HESS J1808-204 \citep[][]{Rowell2012} is consistent with the photon index $2.65 \pm 0.19$ of HESS J1713-381 \citep[][]{Aharonian2008}, which is a TeV PWN produced by the magnetar CXOU J171405.7-381031 \citep[][]{Halpern2010}.  Similarly to SGR 1806-20, CXOU J171405.7-381031 has a spin-down power of $L_\mathrm{sd}\sim4.2\times 10^{34}$ erg s$^{-1}$, a surface magnetic field strength of $B\sim9.6\times10^{14}$ G at the pole \citep[][]{Halpern2010} and hence the power of magnetic field decay  $L_\mathrm{B}\sim10^{36}$ erg s$^{-1}\gtrsim10L_\mathrm{sd}$ \citep[cf. ][]{Zhang2003}.  A major uncertainty of this scenario is that there is no firmly identified PWN in this region, as the leptonic and/or hadronic nature of HESS J1808-204 is currently unclear \citep[][]{Rowell2012}.

At around 2011 January 21, the LAT flux of \emph{Fermi} J1808.2-2029 started an abrupt yet dramatic enhancement with a slight spectral steepening, which lasted for $\lesssim$45 days. It is \emph{unlikely} to be associated with any X-ray outburst of SGR 1806-20. As the enhancement is constrained to occur at energies $\lesssim$400 MeV, we interpret that the enhanced emission is mostly leptonic. Furthermore, according to the 3rd catalog of AGNs detected by the \emph{Fermi} LAT \citep[3LAC; ][]{Ackermann2015}, there is \emph{no} discovered AGN within $3^\circ$ from  the center of \emph{Fermi} J1808.2-2029. Therefore, we speculate the possibility of an independent $\gamma-$ray outburst of SGR 1806-20 occuring at around that epoch.

\section{Summary}

\emph{Fermi} J1808.2-2029/HESS J1808-204 has intense $\gamma-$ray emission, with the spectrum from 300 MeV to 4 TeV well described by a PL model of photon index $\Gamma=2.60 \pm 0.04$.  In terms of the energy budget,  the emission  from 200 MeV to several GeV is easily accounted for by SNR G9.7-0.0 interacting  hadronically with a MC  and/or leptonic particles in stellar winds from Cl* 1806-20. We speculate the possibility that the emission from several GeV to 4 TeV is leptonic and dominated  by a PWN powered by the magnetic-field decay of SGR 1806-20.  Such a `hybrid' scenario is also consistent with the  morphologies we observed: The centroid at the $E_\mathrm{{cut,min}}$ of 200 MeV is almost equidistant from the SNR G9.7-0.0 center and SGR 1806-20/Cl* 1806-20;  in 3-500 GeV, the entire feature is  resolved to be two separated clumps (with a $\sim3\sigma$ significance), whose regions are \emph{respectively} coincident with HESS J1808-204 and SNR G9.7-0.0. 

We confirm that an abrupt yet dramatic enhancement of 60-400 MeV (probably leptonic) LAT flux of \emph{Fermi} J1808.2-2029 occurs from 2011 January 21 to 2011 Mar 7.  Whether it is caused by the magnetar SGR 1806-20 or not remains an open question. Therefore, we strongly encourage the pulsation search of SGR 1806-20 using \emph{Fermi} LAT data within these 45 days.

\section*{Acknowledgement}

This project is supported by the Ministry of Science and Technology of the Republic of China (Taiwan) through grant 103-2628-M-007-003-MY3. P.H.T.T. is supported by the One Hundred Talent Program of the Sun Yat-Sen University and the Fundamental Research Funds for the Central Universities in P. R. China.  C.Y.H. is supported by the National Research Foundation of Korea through grant 2014R1A1A2058590. K.S.C. is supported by a GRF grant of Hong Kong Government under HKU17300814P. P.K.H.Y. thanks X. Hou for useful discussion.

\clearpage

\clearpage

\clearpage

\clearpage

\clearpage

%\footnotetext{\url{http://gcn.gsfc.nasa.gov/gcn3_archive.html}}

\clearpage

\begin{table}
	\caption{The values of the \emph{ln(likelihood)} in 0.2-50 GeV for ``FRONT" data, where the point source 3FGL J1809.2-2016c is replaced with  different morphologies of uniform disks centered at the 1-50 GeV centroid.}
	\begin{center}
		\begin{tabular}{lc}
			\hline \hline
			Radius of extension (deg)	&	 \multicolumn{1}{c}{ln(likelihood)}   \\ 
			\hline
			0\textsuperscript{*}&	7616310.800\\
			0.1&	7616350.215\\
			0.2&	7616360.186\\
			0.3&	7616359.895\\
			0.4&	7616384.921\\
			0.5&	7616429.015\\
			0.6&	7616437.447\\
			0.61&	7616437.961\\
			0.65&	7616438.513\\
			0.7&	7616438.077\\
			0.8&	7616431.815\\
			0.9&	7616387.808\\
			\hline
			\textsuperscript{*}\footnotesize{This corresponds to a point-source model.}
		\end{tabular}
	\end{center}
	\label{morphology}
\end{table}
\clearpage
\begin{table}
	\caption{$\gamma-$ray spectral properties of \emph{Fermi} J1808.2-2029 as observed in 0.2-50 GeV by \emph{Fermi} LAT.}
	\begin{center}
		\begin{tabular}{lc}
			\hline \hline
			
			\multicolumn{2}{c}{PL}   \\ \hline
			$\Gamma$ & 2.44897 $\pm$ 0.0272569  \\ 
			Flux ($10^{-9}$ photons cm$^{-2}$ s$^{-1}$) & 160.924 $\pm$ 4.72334  \\
			TS & 1409.07 \\ \hline
			\multicolumn{2}{c}{PLE}   \\ \hline
			$\Gamma$ & 2.09236 $\pm$ 0.0835688     \\ 
			$E_\mathrm{c}$ (MeV) & 3627.92 $\pm$ 1017.36     \\ 
			Flux ($10^{-9}$ photons cm$^{-2}$ s$^{-1}$) & 157.55 $\pm$ 4.85607     \\ 
			TS & 1450.93     \\ \hline
			\multicolumn{2}{c}{BKPL}   \\ \hline
			$\Gamma_{1}$ & -0.414013 $\pm$ 0.714715     \\ 
			$\Gamma_{2}$ & 2.60408 $\pm$ 0.0442632     \\ 
			$E_\mathrm{b}$ (MeV) & 296.947 $\pm$ 14.6352     \\ 
			Flux ($10^{-9}$ photons cm$^{-2}$ s$^{-1}$) & 156.085 $\pm$ 5.48364     \\ 
			TS & 1476.96     \\ \hline
		\end{tabular}
	\end{center}
	\label{spectral}
\end{table}

\end{document}